\documentclass[
  ,draft            % use draft while you are working on the paper
  ]
  {aipproc}

\layoutstyle{6x9}

\newcommand{\gtrsim}{\mathrel{\hbox{\rlap{\lower.55ex \hbox {$\sim$}}
                   \kern-.3em \raise.4ex \hbox{$>$}}}}
\newcommand{\lesssim}{\mathrel{\hbox{\rlap{\lower.55ex \hbox {$\sim$}}
                   \kern-.3em \raise.4ex \hbox{$<$}}}}

\begin{document}

\title{An absorbed view of a new class of INTEGRAL sources}


\author{E.~Kuulkers}{
  address={ISOC, ESA/ESAC, Urb.\ Villafranca del Castillo,
	   P.O.\ Box 50727, 28080 Madrid, Spain}
}

\begin{abstract}
The European $\gamma$-ray observatory {\em INTEGRAL} has found a group of 
hard X-ray sources which are highly absorbed, i.e., with column densities
higher than about 10$^{23}$\,cm$^{-2}$. Here I give an overview of this class of
{\em INTEGRAL} sources.  The X-ray, as well as
the optical/IR, properties of these sources and their location in the sky
suggest that they belong to the class of high-mass X-ray binaries, some
of them possibly long-period X-ray pulsars. 
The donors in these binaries are most probably giant or supergiant stars.
I suggest that the soft X-ray spectrum below $\sim$5\,keV of IGR\,J16318$-$4848,
as well as in several other X-ray binaries (e.g., XTE\,J0421+56), can be described 
by emission from a compact object which is strongly absorbed by a partionally ionised 
dense envelope.
\end{abstract}

\maketitle

%%%%%%%%%%%%%%%%%%%%%%%%%%%%%%%%%%%%%%%%%%%%
%% MAINMATTER
%%%%%%%%%%%%%%%%%%%%%%%%%%%%%%%%%%%%%%%%%%%%

\section{Introduction}
Hard X-rays (typically $\gtrsim$20\,keV) and $\gamma$-rays are not easily absorbed
by matter and thus are highly penetrating. Such radiation is, therefore, ideal
to probe high-energy emitting sources in dense regions. 
Since its launch in October 2002 {\em INTEGRAL} 
({\em Inte}rnational {\em G}amma-{\em Ra}y {\em L}aboratory; Winkler et al.\ 2003) 
is revealing hard X-ray/soft $\gamma$-ray sources
which were not easily spotted in earlier soft X-ray (typically $\lesssim$10\,keV) observations.
This invited review deals with this apparent new class of X-ray sources which show
intrinsically high absorption along the line of sight, i.e., orders of magnitude
higher than the usual interstellar absorption. 
I will describe their properties, and discuss their nature.

{\em INTEGRAL} carries
on-board four instruments\footnote{For a full description of the instruments, as well
as an account of the first results, I refer to the special A\&A Letters {\em INTEGRAL}
issue 411 (2003).}: 
SPI, a hard X-ray/$\gamma$-ray spectrometer; 
IBIS, two coded mask hard X-ray/$\gamma$-ray imagers, ISGRI and PICsIT;
Jem-X, two identical coded mask X-ray imagers;
OMC, an optical monitor.
The sources described in this paper are mainly found by IBIS/ISGRI 
(Ubertini et al.\ 2003, Lebrun et al.\ 2003).
It has a 8.3$^{\circ}$x8$^{\circ}$ fully-coded field-of-view, while it is 
partially coded out to 29$^{\circ}$x29$^{\circ}$; the
angular resolution is 12'. This makes it ideal for observations in crowded regions.

The large field-of-view of IBIS makes it ideal to map the hard X-ray/$\gamma$-ray sky. 
During the first year of Galactic Plane observations about 120 point sources were
detected down to $\simeq$1\,mCrab (30--100\,keV; Bird et al.\ 2004). 
Among them are previously unknown sources, such as the well-known example
IGR\,J16318$-$4848 (see below). 
About 86\%\ of the Galactic hard X-ray emission up to $\sim$100\,keV 
can be attributed to these high-energy point sources (Lebrun et al.\ 2004).

\section{New, highly absorbed, INTEGRAL sources}

As an illustration of the usefulness of the combination of a large field-of-view and high sensitivity at
hard X-ray/soft $\gamma$-rays, {\em INTEGRAL} discovered its first source,
IGR\,J16318$-$4848, soon after nominal
operation started, on January 29, 2003 during a routine Galactic Plane Scan
(Courvoisier et al.\ 2003). 
Re-analysis of archival {\em ASCA} data revealed that its position coincides with a
highly absorbed ($N_{\rm H}$$\sim$10$^{24}$\,cm$^{-2}$) source with some hint of an Fe emission line
(Murukami et al.\ 2003, Revnitsev et al.\ 2003).
Two weeks after the {\em INTEGRAL} detection, an {\em XMM-Newton} TOO
observation indeed unveiled a variable and heavily absorbed source
(2$\times$10$^{24}$\,cm$^{-2}$), which emitted strong emission lines
(Schartel et al.\ 2003). 
The emission complex could be resolved into three components, with centroid energies of
6.4\,keV, 7.1\,keV and 7.5\,keV.
They are most naturally interpreted as low ionised emission from Fe\,K$\alpha$,
K$\beta$ and Ni\,K$\alpha$ (de Plaa et al.\ 2003, Walter et al.\ 2003, Matt \&\ Guainazzi 2003). 

After IGR\,J16318$-$4848, {\em INTEGRAL} found many more new sources
(hereafter called IGR sources). 
Up to April 2005 more than 50 of these IGR sources have been 
reported.\footnote{See http://isdc.unige.ch/$\sim$rodrigue/html/igrsources.html for
an up-to-date list and further information.} 
Some were identified with already
known sources (e.g., IGR\,J17464$-$3213 = H1743$-$322), but most of them are 
new ones.\footnote{Note, however, that many of the new sources do have catalogued 
{\em ROSAT}, {\em ASCA}, and/or {\em BeppoSAX} soft-energy counterparts.}
About one third of the IGR sources can be classified. Most of them
are either persistent or transient low-mass X-ray binaries (LMXBs) or 
high-mass X-ray binaries (HMXBs). Some of them have been classified as either being
cataclysmic variables (e.g., IGR J17303$-$0601), 
accreting millisecond X-ray pulsar (IGR\,J00291+5934),
AGN (e.g., IGR\,J18027$-$1455), 
or the central source of our Galaxy, Sgr\,A$^{\ast}$ (IGR\,J17456$-$2901). Still, 
about two third of them are unclassified, and
some work lies ahead of us. The distribution of these sources is shown in
Fig.~1. It seems that they are all distributed along the galactic plane, with
concentrations in the direction of the Galactic Center and Galactic arms
(see, e.g., Lutovinov et al.\ 2005b).
One must note, however, that a lot of the {\em INTEGRAL} observations are concentrated
on regions around the Galactic plane and the detection of new (especially transient)
sources may, therefore, be biased towards these regions.

%%%%%%%%%%%%%%%%%%%%%%%%%%%%%%%%%%%%%%%%%%%%
\begin{figure}
  \includegraphics[height=.5\textheight,angle=-90]{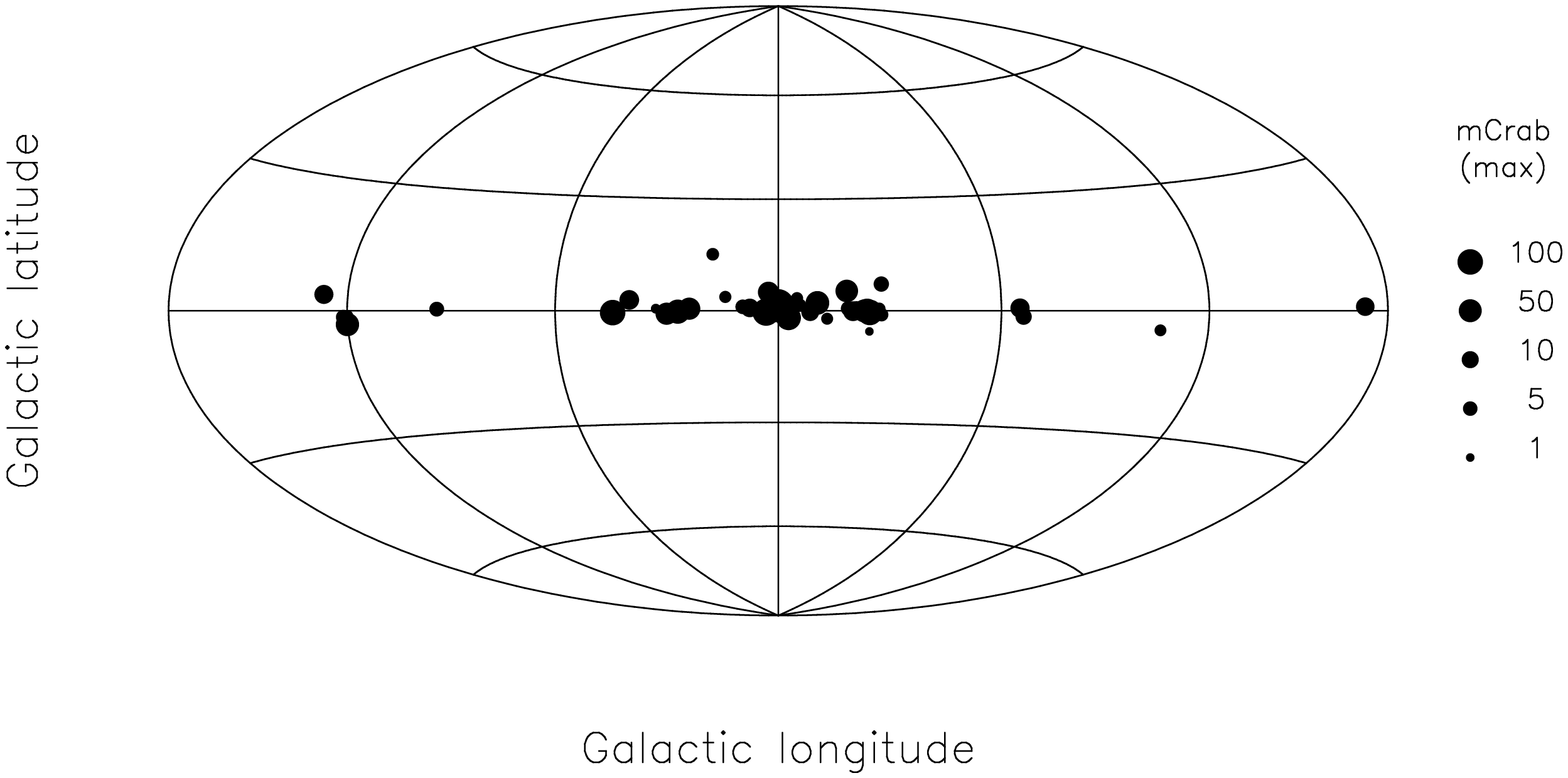}
\end{figure}
\begin{figure}
  \includegraphics[height=.5\textheight,angle=-90]{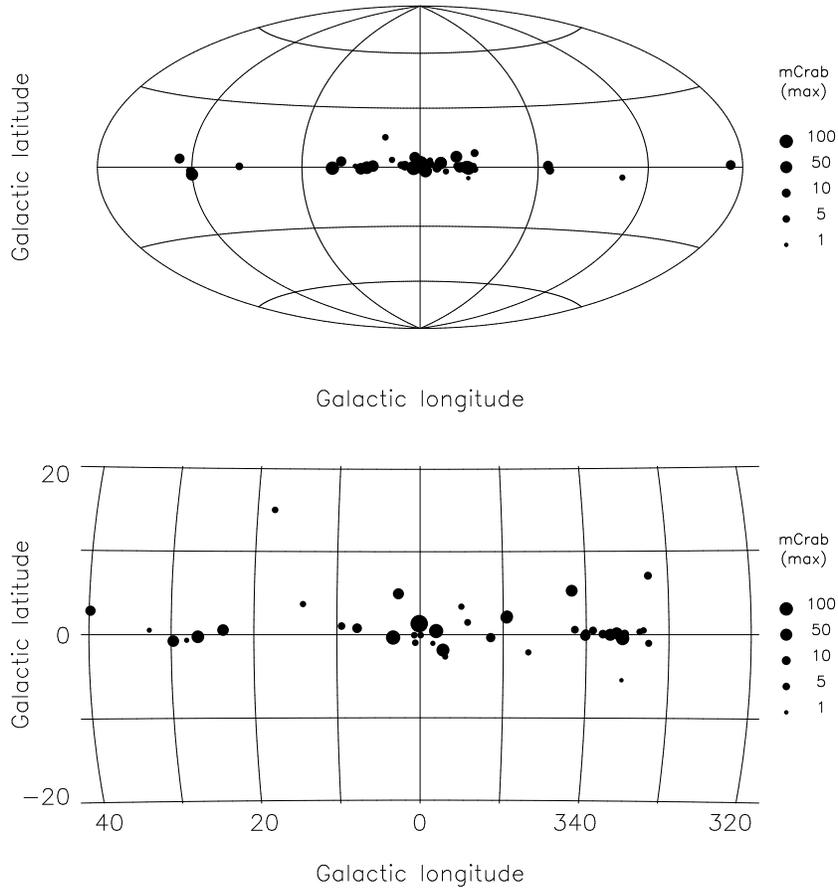}
  \caption{Galactic distribution of IGR sources. The size of the symbol $\bullet$ indicates
the maximum observed X-ray intensity in units of mCrab in the 20-40\,keV or 20-60\,keV band.
Flux values were mostly taken from those reported in the Astronomer's Telegrams and
IAU Circulars.}
\end{figure}
%%%%%%%%%%%%%%%%%%%%%%%%%%%%%%%%%%%%%%%%%%%%

Of the (up to now well-studied) IGR sources, ten of them show very strong absorption
(N$_{\rm H}$$\gtrsim$10$^{23}$\,cm$^{-2}$), i.e., one to two orders of magnitude higher
than the Galactic value of around 10$^{22}$\,cm$^{-2}$.
It is this class of sources which I concentrate on for the rest of this paper,
and I will refer to them as highly absorbed IGR sources. 
They are listed in Table 1; one thing which immediately catches the eye is that 
all but one (IGR\,J19140+0951) are in the direction of the Norma-arm tangent region.
I will come back to this later.

\begin{table}
\begin{tabular}{lccccc}
\hline
\tablehead{1}{c}{b}{IGR source\\ }
& \tablehead{1}{c}{b}{N$_{\rm H}$\\(10$^{23}$\,cm$^{-2}$)}
& \tablehead{1}{c}{b}{$\Gamma$\\ }
& \tablehead{1}{c}{b}{E$_{\rm cut}$\\(keV)}
& \tablehead{1}{c}{b}{P$_{\rm pulse}$\\(min)}
& \tablehead{1}{c}{b}{references$^a$\\ } \\
\hline
IGR\,J16195$-$4945 & $\simeq$1  & $\sim$0.6 &               & & [1] \\
IGR\,J16318$-$4848 & $\simeq$20 & 1.7--2.1  & $\simeq$15    & & [2],[3] \\
IGR\,J16320$-$4751 & $\simeq$2  & 0.5--1.1  & $\simeq$12    & $\simeq$22  & [4],[5],[6] \\
IGR\,J16358$-$4726 & $\simeq$4  & $\sim$0.7 & $\simeq$16    & $\simeq$100 & [7],[8] \\
IGR\,J16393$-$4643 & $\simeq$6  & $\sim$1.3 & $\simeq$11    & $\simeq$15  & [8],[9],[10] \\
IGR\,J16418$-$4532 & $\gtrsim$1 &           &               & & [11] \\
IGR\,J16493$-$4348 & $\simeq$1  & $\sim$1.4 &               & & [12] \\
IGR\,J16465$-$4507 & $\simeq$7  & $\sim$1   & $\simeq$32    & $\simeq$4 & [8] \\
IGR\,J16479$-$4514 & $\simeq$1  & $\sim$1.4 &               & & [8] \\
IGR\,J19140+0951   & $\simeq$1$^b$ & $\simeq$1.6 &          & & [13] \\ 
\hline
\multicolumn{6}{l}{\footnotesize $^a$ References: [1] Sidoli et al.\ (2005), [2] Walter et al.\ (2003), [3] Matt \& } \\
\multicolumn{6}{l}{\footnotesize Guainazzi (2003), [4] Rodriguez et al.\ (2003), [5] Foschini et al.\ (2004),}\\
\multicolumn{6}{l}{\footnotesize [6] Lutovinov et al.\ (2005a), [7] Patel et al.\ (2004), [8] Lutovinov et al.\ (2005b),} \\ 
\multicolumn{6}{l}{\footnotesize [9] Combi et al.\ (2004), [10] Walter (2005), [11] Walter et al.\ (2004),} \\
\multicolumn{6}{l}{\footnotesize [12] Markwardt et al.\ (2005), [13] Rodriguez et al.\ (2005a),} \\
\multicolumn{6}{l}{\footnotesize $^b$ This is an observed maximum value which is only reached occassionally.} \\
\end{tabular}
\caption{Properties of highly absorbed (N$_{\rm H}$$\gtrsim$10$^{23}$\,cm$^{-2}$) {\em INTEGRAL} sources}
\end{table}

Some of the highly absorbed IGR sources seem to be more or less persistent 
(such as IGR\,J16318$-$4848; see, e.g., Matt et al.\ 2005); 
some of them are clearly transient (e.g., IGR\,J16358$-$4726:
Patel et al.\ 2004; IGR\,J16465$-$4507: Lutovinov et al.\ 2005b). 
The highly absorbed IGR sources vary in brightness on time scales of minutes to hours, as
well as from observation to observation, both at soft and hard X-ray energies 
(e.g., IGR\,J16318$-$4848: Walter et al.\ 2003, Matt \&\ Guainazzi 2003,
Matt et al.\ 2005; IGR\,J16320$-$4751: Rodriguez et al.\ 2003, Foschini et al.\ 2004). 
In IGR\,J16318$-$4848 the line emission also varies on time scales of
$\sim$15\,min and longer (Matt \&\ Guainazzi 2003).
Up to now, four of the (well-studied) highly absorbed IGR sources have been seen to also vary on a regular time scale
(between $\simeq$4 to $\simeq$100\,min, see Table 1).
IGR\,J16358$-$4726, for example, displayed a strong period flux modulation
with a peak-to-peak pulse fraction of $\sim$70\%.
These have been interpreted as neutron star pulse periods.
The absorption column is also seen to vary, from observation to observation
(e.g., IGR\,J16318$-$4848: Revnivtsev 2003; IGR\,J16320$-$4751: Rodriguez et al.\ 2003;
IGR\,J19140+0951: Rodriguez et al.\ 2005a).

In Table 1, I give the parameters for the (cut-off) power-law model\footnote{The
cut-off power-law model is given by 
$e^{-{\rm N}_{\rm H}\sigma(E)}{\rm A}_{\rm pl} {\rm E}^{\Gamma} e^{{\rm E}/{\rm E}_{\rm cut}}$, 
where N$_{\rm H}$ is the absorption column density, $\sigma(E)$ the absorption cross-section,
A$_{\rm pl}$ the power-law normalization, $\Gamma$ the power-law index,
and E$_{\rm cut}$ the cut-off energy.}, which is usually used to fit the soft X-ray 
({\em ASCA}, {\em XMM-Newton}, {\em Chandra}) spectral data, most of the 
time, in combination with {\em INTEGRAL} spectral data. The spectra are hard 
($\Gamma$$\lesssim$2), and show evidence for high-energy cut-off values. However, 
one must note here that the soft-energy spectra were not taken simultaneously
with the hard X-ray spectra. Since the sources are (highly) variable, 
both at soft and hard X-ray energies, one must be cautious with the spectral fitting results
(see, e.g., Walter et al.\ 2003).
Of the highly absorbed IGR sources, only IGR\,J16138$-$4848 shows very strong emission lines
(see above). The others only show weak (or undetectable) line emission.

\subsection{X-ray and optical/IR emission}

As shown in the previous Section, the highly absorbed IGR sources have similar 
X-ray properties: hard X-ray 
(cut-off; E$_{\rm cut}$$\gtrsim$10\,keV) power-law ($\Gamma$$\lesssim$2)
emission and some, or most of the time, strong (N$_{\rm H}$$\gtrsim$10$^{23}$)
and variable absorption. Four of them have been found to show (long-period; $\gtrsim$4\,min)
`pulse' periods.
These properties are rather typical for accreting X-ray pulsars in HMXBs
(see, e.g., White et al.\ 1983), suggesting that the highly absorbed IGR sources
are HMXBs. 
The highly absorbed IGR sources are located in the direction of the Norma spiral-arm tangent, except for
IGR\,J19140+0951. The latter is located in the direction of the Sagittarius spiral-arm
tangent; various HMXB X-ray pulsars are located in that direction, as well 
as other IGR sources (see, e.g., Molkov et al.\ 2004). These regions have an enhanced
concentration of young massive stars (e.g., Grimm et al.\ 2002), and supports the
HMXB hypothesis for the highly absorbed IGR sources. 
If they lie indeed in the above-mentioned spiral arms, their
unabsorbed $\sim$2--100\,keV X-ray luminosity would be around 10$^{35}$ to a few times 
10$^{36}$\,erg\,s$^{-1}$, also very typical for HMXBs.
Note that various other of the less absorbed IGR sources are classified as bonafide
HMXB/Be-X-ray transients (see, e.g., Negueruela 2005).

There are a few well-known HMXBs which show strong (N$_{\rm H}$$\gtrsim$10$^{23}$\,cm$^{-2}$)
and variable absorption, as well as strong Fe\,K$\alpha$ emission and X-ray pulsations:
GX\,301$-$2 (Swank et al.\ 1976, Endo et al.\ 2002), Vela~X-1 (Haberl \&\ White 1990),
XTE\,J0421+56 (CI\,Cam; Boirin et al.\ 2002). The latter is an atypical Be/X-ray 
binary (see below). Note that GX\,1+4, not an HMXB but a rare type of symbiotic LMXB
(see below), showed also similar spectra during an extended low state 
(Naik et al.\ 2005). Other known HMXBs exist which show similarly hard and strongly absorbed
X-ray spectra (e.g., EXO\,1722$-$363: Tawara et al.\ 1989, Takeuchi et al.\ 1990, see also
Walter 2005; 4U\,1909+07: Levine et al.\ 2004).
The fact that the absorption is seen to vary in the HMXBs and the highly absorbed IGR sources, 
suggests it is intrinsic. For example, in GX\,302$-$1 the absorption varies between
3$\times$10$^{23}$ and 2$\times$10$^{24}$\,cm$^{-2}$, and is indeed connected to its 41~day orbit 
(e.g., White \&\ Swank 1984, Endo et al.\ 2002).

Long X-ray `pulse' periods 
have been found previously in HMXBs (but in those which show only moderate 
X-ray absorption), such as SAX\,J2239.3+6116 (21\,min: in 't Zand et al.\ 2001), 4U\,2206+54
($\simeq$60\,min; Masetti et al.\ 2004) and 4U\,0114+650 (2.7\,hr: Finley et al.\ 1992).
4U\,2206+54 and 4U\,0114+650 shows noticeable similar pulsation profiles as those seen
in the highly absorbed IGR sources. Most of the X-ray pulsar HMXBs with pulse period 
longer than typically a few minutes are considered to be pulsars fed by a stellar wind (e.g., Nagase 1989).
In the classical Corbet (1986) diagram the highly absorbed IGR sources
either fall in the group of supergiant systems or the long orbital period ($\gtrsim$30\,days)
Be/X-ray transients. So far, only IGR\,J19140+0951 has a reported orbital period
(P$_{\rm orb}$$\simeq$13.55\,days, Corbet et al.\ 2004).\footnote{Ofcourse, it is assumed 
here that the highly absorbed IGR sources are binaries; this, however, remains to be verified. 
In this respect it is interesting to note that no convincing orbital period has been 
reported for XTE\,J0421+56 either (see, e.g., Hynes et al.\ 2002, for a discussion).
Note also, that the HMXB interpretation was questioned by Patel et al.\ (2004), when discussing the 
periodic X-ray variations in IGR\,J16358$-$4726. They argued that even extremely small
amounts of accretion can spin up the star to shorter periods than those observed.
The only way out might be a pulsar in a Be/X-ray binary, which is able to 
spin down due to the propellor effect during the long quiescence period in between 
outbursts. Interestingly, the pulsations seen in 4U\,0114+650 have been interpreted 
as pulsations from its early B star donor (Finley et al.\ 1992).}
If the highly absorbed IGR sources indeed contain (slow) pulsars, their compact object
is evidently a neutron star. However, if this interpretation is not correct,
a black hole can not be excluded either (this is especially true for those
IGR sources with no identified pulsations).

The presence of strong absorption in the X-ray domain shows that the 
compact object must be embedded in a dense circumstellar envelope, originating from a
dense stellar wind from the donor.
This (relatively cold) envelope also serves as the source of the fluorescent emission, especially
in IGR\,J16318$-$4848 (e.g., Walter et al.\ 2003; Matt \&\ Guainazzi 2003; Revnivtsev et al.\ 2003).

About 70\%\ of the mass donors in HMXB are classical Be stars, the rest are blue
supergiants. A big fraction of the accretion powered X-ray pulsars in the 
Be-systems are transients. Be stars show rich emission line their spectra. 
There is, however, a subclass of objects which also show forbidden lines, as well as
a near-IR excess. These are the B[e] stars;
they include many objects of different types and evolutionary status
(e.g., Lamers et al.\ 1998). 
The typical mass-loss rates in these stars are \.M$\gtrsim$10$^{-6}$\,M$_{\odot}$\,yr$^{-1}$.

A few of the highly absorbed IGR sources have been identified in the optical and/or
IR.  The IR spectra of IGR\,J16318$-$4848 are rich in emission lines, i.e., 
various order H-lines, He\,{\sc I} and {\sc II}, low excitation permitted lines,
as well as forbidden iron lines; some lines show P-Cygni profiles (Filliatre \&\ Chaty 2004). 
Many of these IR spectral lines can be identified in XTE\,J0421+56 too (see Clark et al.\ 1999).
For both XTE\,J0421+56 (e.g., Clark et al.\ 1999, Hynes et al.\ 2002) and 
IGR\,J16318$-$4848 (Filliatre \&\ Chaty 2004) it has been suggested that they have a supergiant B[e] donor 
present in a dense and absorbing circumstellar environment.
Comparable IR spectra are also seen in GX\,1+4. Its donor is, however, a cool giant star,
and the IR spectra show in addition late-type features,
such as CO-bands (Clark et al.\ 1999). These features are not seen in either 
XTE\,J0421+56 or IGR\,J16318$-$4848.
The donor in IGR\,J16465$-$4507 has also been suggested to be an (early) supergiant
(Smith 2004), whereas in IGR\,J16320$-$4751 it is either a cool giant or supergiant
(Rodriguez et al.\ 2003). Early-type stars are also found in the error circles for
some of the other highly IR absorbed sources.

The near-IR excess in B[e] stars points to the presence of hot circumstellar dust.
Both in IGR\,J16320$-$4751 (Rodriguez et al.\ 2003) and IGR\,J16465$-$4507
(Smith 2004) there is evidence for such a near-IR excess. 
The IR spectra of IGR\,J16318$-$4848
suggest a similar configuration (Filliatre \&\ Chaty 2004).

The column density derived from the optical extinction is found to be
one to two orders of magnitude less than that derived from the X-ray measurements
(IGR\,J16318$-$4848: Walter et al.\ 2003, Filliatre \&\ Chaty 2004;
IGR\,J16320$-$4751: Rodriguez et al.\ 2003; IGR\,J16465$-$4507: Smith 2004).
This suggests that the dense circumstellar envelope must be rather compact
and concentrated towards the compact object (e.g., Revnivtsev et al.\ 2003).

\subsection{Soft X-ray excess}

Soft X-ray excess emission between 0.3 and 5\,keV has been reported for IGR\,J16318$-$4848
(when fitting the observed X-ray spectrum with a power-law spectrum to standard absorption column,
see footnote on page 4). It seems to be consisting of two parts, one above and one below
$\sim$2\,keV.  Partial covering could account for the excess between $\sim$2--5\,keV.
If the covering fraction is less than 1, part of the X-ray illuminated surface should be directly visible,
producing a Compton reflection component (Matt \&\ Guainazzi 2003, Matt et al.\ 2005, and references therein).
It is not clear whether such a component is present or not 
(Walter et al.\ 2003, Matt \&\ Guainazzi 2003). The excess between 0.3--2\,keV could not be easily explained
(Matt \&\ Guainazzi 2003). Hints of a soft X-ray excess are seen as well in IGR\,J16320$-$4751
(Rodriguez et al.\ 2005b).

There is evidence for an (independent) soft component in the 
X-ray spectrum of XTE\,J0421+56
in outburst, subsequent decay, and quiescence, below a few keV (Boirin et al.\ 2002, 
Ishida et al.\ 2004, and references therein).
A similar feature is seen in the soft X-ray spectrum of 
CH\,Cyg, a symbiotic star containing a white dwarf
(CH\,Cyg; Ezuka et al.\ 1998). Modeling the parts below and above $\sim$2\,keV
as separate emission components (Ezuka et al.\ 1998; Ishida et al.\ 2004), or modeling the 
emission with a partial covering absorption model (e.g., Boirin et al.\ 2002)
seemed to work fine.
Because of the similarity with CH\,Cyg, it was proposed that XTE\,J0421+56
contains a white dwarf (Ishida et al.\ 2004); this is, however, hard to reconcile 
with the observed X-ray properties of XTE\,J0421+56 (e.g., Hynes et al.\ 2002, and 
references therein) and IGR\,J16318$-$4848.
It is interesting to note that soft X-ray spectra during X-ray dips in the light curves of 
LMXBs and HMXBs also show an excess
in emission below typically 4\,keV (see, e.g., Kuulkers et al.\ 1998, and references therein).
Such dips are thought to be due to strong absorption
(up to a few 10$^{23}$\,cm$^{-2}$) of emission from the inner parts of the
accretion disk and compact object by the cooler outer parts of the accretion disk.

Wheatley (2001) showed that the soft X-ray spectrum of CH\,Cyg can be solely described 
by emission from the white dwarf which is strongly absorbed by a partionally ionised wind 
from the red giant. Recently, a partionally ionised absorber has been proposed as well
to explain the soft X-ray spectra during the dips in LMXBs
(Boirin et al.\ 2005).
Since the soft spectral properties of XTE\,J0421+56 and IGR\,16318$-$4848 are very similar
to CH\,Cyg, it is logical to suggest such a model for these sources as well. 
The strong emission from the compact object is able to ionize the immediate environment.
Such a ionised region can be responsible for, e.g., the observed IR He\,{\sc II} emission
in XTE\,J0421+56 and IGR\,16318$-$4848.
If this works,
than there may not be a need for a Compton reflection component and one can conclude that the
whole X-ray emitting region is covered by the circumstellar material.

\section{Summary}

The above described X-ray and optical/IR properties
suggests that the highly absorbed IGR sources are HMXBs containing 
either a neutron star or black hole in orbit around a (super)giant
donor. The stellar wind accreting
onto the compact object could form a dense envelope in which absorption,
fluoresence and ionization takes place. 
This circumstellar envelope does not seem to cover much of the (super)giant
donor.
Because of the wavelength window {\em INTEGRAL} is able to observe, we are
now starting to find more of this previously poorly known class of sources.
Indeed, thanks to INTEGRAL ``we can see clearly now ...'' (White 2004).

\begin{theacknowledgments}
I am indebted to the conference organizers and editors, who allowed
me to review these intriguing class of sources driven by results from {\em INTEGRAL}.
I thank Deepto Chakrabarty for drawing my attention to the
IR observations of GX\,1+4, and J\'er\^ome Rodriguez for discussions on an earlier draft
of this paper.
\end{theacknowledgments}

\bibliographystyle{aipproc}   % if natbib is available

\end{document}